\documentclass[a4paper,11pt]{article}
\usepackage{amsmath}
\usepackage{mathrsfs}
\usepackage{geometry}
\usepackage{cancel}
\usepackage{multicol}
\usepackage{tikz}
\usepackage{breqn}
\usepackage{braket}
\usepackage{soul}
\usepackage{tikz-feynman}
\usepackage{graphicx}
\usepackage[english]{babel}
\usepackage{subcaption}
\usepackage{float}
\usepackage{lipsum}
\usepackage{hyperref}
\usepackage{float}
\usepackage[normalem]{ulem}
\usepackage{multirow}
\usepackage[sort&compress]{natbib}
\hypersetup{
    colorlinks=true,
    citecolor=cyan,
    linkcolor=blue,
    urlcolor=blue}
\usepackage{amssymb}
\usepackage{pifont}
\usepackage{authblk}
\def\beq{\begin{equation}}
\def\eeq{\end{equation}}
\def\barr{\begin{array}}
\def\earr{\end{array}}

\newcommand{\be}{\begin{equation}}
\newcommand{\ee}{\end{equation}}
\newcommand{\bea}{\begin{eqnarray}}
\newcommand{\eea}{\end{eqnarray}}

\newcommand{\bi}{\begin{itemize}}
	\newcommand{\ei}{\end{itemize}}

\evensidemargin=\oddsidemargin
\title{On the role of cosmological constant in modeling hadrons}
\author{Mathew Thomas Arun\footnote{mathewthomas@iisertvm.ac.in}}
\author{Nabeel Thahir\footnote{nabeelthahir20@alumni.iisertvm.ac.in }}
\affil{School of Physics, Indian Institute of Science Education and Research, Thiruvananthapuram-695551, Kerala, India}
\begin{document}
  \maketitle

  \begin{abstract}
Einstein’s gravity with a cosmological constant $\Lambda$ in four dimensions can be reformulated as a $\lambda \phi^4$ theory characterized solely by the dimensionless coupling $\lambda \propto G_N \Lambda$ ($G_N$ being Newton’s constant). The quantum triviality of this theory drives $\lambda \to 0$, and a deviation from this behavior could be generated by matter couplings. Here, we study the significance of this conformal symmetry and its breaking in modeling non-perturbative QCD. The hadron spectra and correlation functions are studied holographically in an $AdS_5$ geometry with induced cosmological constants on four-dimensional hypersurface. Our analysis shows that the experimentally measured spectra of the $\rho$ and $a_1$ mesons, including their excitations and decay constants, favour a non-vanishing induced cosmological constant in both hard-wall and soft-wall models. Although this behavior is not as sharp in the soft-wall model as in the hard-wall model, it remains consistent. Furthermore, we show that the correction to the Gell–Mann–Oakes–Renner relation has an inverse dependence on the induced cosmological constant, underscoring its significance in holographic descriptions of low-energy QCD.

  \end{abstract}

\section{Introduction} 
Conformal symmetry and its breaking are well-studied features across a wide range of systems, including particle physics, condensed matter, and cosmology. Within the context of gravity, an interesting realization of this idea arises in four dimensions, where Einstein’s theory of gravity with a cosmological constant ($\Lambda$) can be recast in a scale-invariant form~\cite{deAlfaro:1978dz,Polyakov:2000fk,Jackiw:2005yc,Cadoni:2006ww},
\begin{equation}
S = \frac{1}{16\pi G_N} \int d^4x \, \sqrt{-g} \, ( R - 2\Lambda )
 \, \, \longrightarrow \, \,
\frac{3}{4\pi} \int d^4x \left[ \frac{1}{2} (\partial \phi)^2 - \frac{\lambda}{4!} \phi^4 \right],
\end{equation}
upon fixing the diffeomorphism invariance of the metric to a conformally flat space-time,
\begin{equation}
g_{\mu\nu} = \phi^2 \eta_{\mu\nu},
\end{equation}
and redefining the field as $\phi \to \phi / \sqrt{G_N}$. Here, $\eta_{\mu\nu}$ denotes the metric of four-dimensional flat space-time. The resulting action depends only on the dimensionless coupling constant $\lambda = 4 G_N \Lambda$.
In the absence of matter, and restricting to conformally flat perturbations of the metric, Einstein’s gravity with a cosmological constant can thus be viewed as a four-dimensional conformal field theory. This conformal symmetry, however, is generally broken by the inclusion of matter through an additional term $S_{\text{matter}}$,
\begin{eqnarray}
S &=& \frac{3}{8\pi} \int d^4x \left[ (\partial \phi)^2 - \frac{\lambda}{4!} \phi^4 \right] + S_{\text{matter}} .
\label{eq:actionmatter}
\end{eqnarray}

Though the triviality of $\lambda \phi^4$ theory drives the constant $\lambda \to 0$, a small value can be generated via conformal symmetry breaking terms from $S_{matter}$~\cite{Cadoni:2006ww}. This framework has been explored in the literature to address the cosmological constant problem~\cite{Polyakov:2000fk,Jackiw:2005yc,Cadoni:2006ww,Hinterbichler:2012mv}. From the perspective of non-perturbative QCD, the analogous question concerns how conformal symmetry breaking of the theory in Eq.\ref{eq:actionmatter}, or the smallness of the ``cosmological constant" of the mathematical model describing hadrons, influences hadron spectra and correlation functions in the strongly coupled regime~\cite{Mamani:2019mgu}. In this context, ``cosmological constant" does not refer to the uniform energy density in the universe, as discussed in previous literature, rather, it is a parameter that quantifies the extent of conformal symmetry breaking. 

Here, our aim is to understand the influence of this parameter in conformal symmetry breaking by holographically mapping the problem to a $AdS_5$ geometry with non-vanishing cosmological constant induced on the embedded four-dimensional hypersurfaces. Understanding low-energy QCD has been one of the most important areas in high-energy physics for a long time. Despite the vast amount of data from hadronic measurements, QCD has eluded a complete solution due to its non-perturbative nature at low energies. Since the advent of the AdS/CFT correspondence~\cite{osti_361743}, the area has received immense attention, owing to the successes of the Sakai-Sugimoto model\cite{10.1143/PTP.113.843, 10.1143/PTP.114.1083} with D-brane configurations. This model was later generalized to include additional branes/technicolor sectors to explain the pion mass \cite{OdedMintakevich_2009, PhysRevD.77.115002}. 

Alongside these string theoretic models, there have been independent efforts to describe the 5-dimensional dual gravity side using low-energy observations. These 'bottom-up' field-theoretic approaches \cite{HadronModel, LeandroDaRold_2006} are characterized by solvable perturbative calculations of fields and their Kaluza-Klein modes in the bulk, and their matching with hadronic masses and couplings on the IR brane. The remarkable success of this 'bottom-up' approach has warranted a detailed study of the setup with various deformations. Unfortunately, in this method, the gravity in the 5-dimensional bulk remains an open question. Recent methods of AdS/DL correspondence \cite{PhysRevD.98.046019, PhysRevD.98.106014} aim to answer such questions partially. 

The gauge/gravity duality maps the underlying field theory into a classical gravitational theory which is asymptotically $AdS_5$. This geometry arises from a five-dimensional warped space-time with branes located at the two orbifold fixed points, carrying equal but opposite brane tensions~\cite{RSmodel, PhysRevLett.83.4690}. This construction, usually, assumes vanishing induced cosmological constant, ensuring a flat Minkowski metric on the four-dimensional hypersurfaces. In this work, our objective is to understand the holographic QCD framework with non-vanishing induced cosmological constant, particularly in generating the parameter $\lambda$ within the conformal symmetry breaking terms of Eq.~\ref{eq:actionmatter}. This breaking is captured in the Gell-Mann--Oakes--Renner (GOR) \cite{PhysRev.175.2195}. In the limit of vanishing quark masses, the pion would be massless, and thus realized as Goldstone bosons of chiral symmetry breaking. However, in $AdS_5$ with a non-vanishing induced cosmological constant ($\omega$), we observe that the corrections to this relation has an inverse dependence on $\omega$ and is expressed as,
\[
m_\pi^2 f_\pi^2 = - (m_u + m_d) \langle \bar{q} q \rangle + \mathcal{F}(z_m,\omega) \ ,
\]
where, $z_m$ is the location of the $IR$ brane in the extra-dimension.
Here, we also show through a $\chi^2$ analysis that the meson spectrum and decay constants, fit better with the predictions of $AdS_5/QCD$ with a non-vanishing induced cosmological constant than compared to the previous studies~\cite{HadronModel, LeandroDaRold_2006, PhysRevD.74.086004, PhysRevD.78.025001}. Thus, modeling hadrons in a broken conformal sector with non-vanishing $\lambda$ as in Eq.\ref{eq:actionmatter} fits better with the experiments.

The article is organized as follows. In section \ref{inducedccAdS}, we start with briefly introducing the induced cosmological constant in $AdS_5$ space-time. We then consider a hard wall $AdS_5/QCD$ model with the 5D chiral Lagrangian and its symmetry breaking in section \ref{bentbranehadrons}. Further, we examine the dilaton model (soft wall) and predict meson resonances in the presence of an induced cosmological constant in section \ref{dilaton}. We also provide the numerical analysis and compare our model with the standard $AdS_5$ prediction of meson masses and decay constants, and summarize our results and findings in section \ref{summary}.

\section{\texorpdfstring{$AdS_5$}{ADS5} with induced cosmological constant}
\label{inducedccAdS}
Consider a five-dimensional warped compactified extra-dimensional space-time orbifolded on $Z_2$ to break the 5D symmetry $M^{1,4} \xrightarrow{}M^{1,3} \times S^{1}/Z_{2}$~\cite{RSmodel}. The metric for this geometry is defined by,
\begin{equation}
    ds^2 = e^{-2A(y)} g_{\mu\nu} dx^\mu dx^\nu -  dy^2
    \label{eqn1}
\end{equation}
where the compact space is represented by the coordinate $y(r,\theta)$ with $r$ being the moduli and $\theta \in [0,\pi]$. The general warp factor is represented by $e^{-A(y)}$, with the hypersurfaces at $y = 0$ and $y = r \pi$ identified as the UV and IR branes, respectively. The total bulk-brane action for the 5-dimensional space-time is, thus,
\begin{equation}
\label{eqn2}
S = \int d^5x \sqrt{-G}(M^3 R - \Lambda) + \int d^4x \sqrt{-g_i} \mathcal{V}_i \ .
\end{equation}
Here, $\Lambda$ represents the bulk cosmological constant, $R$ is the bulk five-dimensional Ricci scalar, and $\mathcal{V}_{i}$ is the brane tension on the $i^{th}$ brane. Also, $G$ denotes the determinant of the five-dimensional bulk metric, while $g_i$ denotes the determinant of the metric induced on the $i{}^{th}$ brane. For the rest of the paper, $g_{\mu \nu}$ represents the four-dimensional metric that we get after orbifolding. Upon varying with respect to the bulk metric, the resulting Einstein equations become,
\begin{equation}
    {}^4 G_{\mu\nu} - g_{\mu\nu} e^{-2A} \left[ -6A'{}^2 + 3A'' \right] = -\frac{\Lambda}{2M^3} g_{\mu\nu} e^{-2A} \ ,
    \label{eqn3}
\end{equation}
\begin{equation}
    -\frac{1}{2} e^{2A} \,{}^4R + 6A'{}^{2} = -\frac{\Lambda}{2M^3} \ ,
    \label{eqn4}
\end{equation}
with the boundary conditions,
\begin{equation}
    \left[ A'(y) \right]_i = \frac{\epsilon_i}{12M^3} \mathcal{V}_i \ ,
    \label{eqn5}
\end{equation}
where $\epsilon_{UV} =- \epsilon_{IR}=1$ for the case of Randall-Sundrum model. Also, ${}^4G_{\mu\nu}$ and ${}^4R$ in the above relations are the four-dimensional Einstein tensor and Ricci scalar, respectively, in the orbifolded geometry defined with the four-dimensional metric,  $g_{\mu \nu}$, of the $M^{1,3}$ space-time. 

Using separation of variables, we can write Eq. \ref {eqn3} as,
\begin{equation}
    {}^4 G_{\mu\nu} = -\Big(-e^{-2A} \left[ -6A'{}^2 + 3A'' \right]+\frac{\Lambda}{2M^3} e^{-2A}\Big)g_{\mu\nu}=-\Omega g_{\mu\nu} \ ,
    \label{eqn6}
\end{equation}
where $\Omega$ is a constant of separation. From a four-dimensional perspective, this equation then suggests that the constant $\Omega$ is an \textit{induced cosmological constant} with the relation to the bulk cosmological constant as,
\begin{equation}
    e^{-2A} \left[ -6A'{}^2 + 3A'' - \frac{\Lambda}{2M^3} \right] = -\Omega \ .
    \label{eqn7}
\end{equation}
Assuming $\Omega < 0$, this relation yields the solution for the warp factor~\cite{Das_2008} as, 
\begin{equation}
    e^{-A(y)} = \omega \cosh \left( \ln \frac{\omega}{c_1} + k y \right) \ ,
    \label{eqn8}
\end{equation}
where $c_1$ is the constant of integration $\omega^2 \equiv  -\Omega/3k^2 \geq 0$ is the dimensionless quantity that represents the induced cosmological constant. For consistency of the semi-classical treatment followed in this paper, $\omega$ should be $\leq 0.1$. 

Therefore, the line element for the five-dimensional space-time with induced cosmological constant becomes,
\begin{equation}
    ds^2 = \omega^2 \cosh^2 \left( \ln \frac{\omega}{c_1} + k y \right) g_{\mu\nu} dx^\mu dx^\nu -   dy^2
    \label{eqn9}
\end{equation}
which, when written in conformal coordinates, will take the form,
\begin{equation}
    ds^2 =  \omega ^2\csc^2 (k w z) \left(-dz^2 + dx^\mu dx_\mu \right), \quad 0 < z \leq z_m.
    \label{eqn10}
\end{equation}
In this language, the infrared brane is located at $z = z_m$. In addition, a UV cutoff can be provided by setting the boundary to $z_0 = \epsilon$ with $\epsilon \to 0$. In the rest of the article, we frequently use this cutoff, but we always imply the limit of  $z_0 \xrightarrow{} 0$. 

Unlike the Randall-Sundrum model with vanishing induced cosmological constant, here, the validity of the metric in Eq.\ref{eqn10} requires that $z_m \leq \frac{\pi}{2 k \omega}$, leading to a natural cutoff for this geometry. In the limit $\omega \xrightarrow[]{}0$, the line element Eq.\ref{eqn10} matches with the regular Randall-Sundrum model with the anti-de Sitter ($AdS_5$) metric,
\begin{equation}
    ds^2 =  \frac{1}{(kz)^2} \left(-dz^2 + dx^\mu dx_\mu \right), \quad 0 < z \leq z_m \leq \frac{\pi}{2 k \omega} \ . 
    \label{ads}
\end{equation}
Interestingly, the metric in Eq.\ref{eqn10} does revert to the $AdS_5$ Eq.\ref{ads} in the $z \to 0$ limit, while it deviates in the large $z$ limit. In this article, we explore this feature to understand the influence of the induced cosmological constant on the $AdS_5/QCD$ model for hadrons.
\section{5D action and chiral symmetry breaking (Hard Wall)}
\label{bentbranehadrons}
Instead of a 'top-down' approach of Super Yang-Mills theory to obtain QCD~\cite{CsabaCsaki_1999, GIRARDELLO2000451,PhysRevD.69.066007,HenriqueBoschi-Filho_2003}, we start from QCD and attempt to construct its five-dimensional (5D) holographic dual~\cite{HadronModel, LeandroDaRold_2006, PhysRevD.74.086004, PhysRevD.78.025001}, considering the effects of an induced cosmological constant, with metric given in Eq.\ref{eqn10}. 

The $AdS/CFT$ correspondence is an equivalence between the generating functional, with the sources set to the value of the five-dimensional bulk field at the UV brane, and the effective five-dimensional theory. Since, in this article, we are interested in introducing the model and its implications in the simplest QCD systems, we choose vector and axial-vector meson masses and their resonances, which are studied in detail. The chiral action in the bulk, satisfying $SU(N_f)_L\times SU(N_f)_R$ flavor symmetry, is given as,
\begin{equation}
    S = \int d^5x \sqrt{G} \, \text{Tr} \left\{ |DX|^2 - M_5^2 |X|^2 - \frac{1}{4g_5^2} (F_{MN \ L}^2 + F_{MN \ R}^2) \right\}
    \label{eq12}
\end{equation}
where $D_\mu X = \partial_\mu X - i A_{\mu L}X + i X A_{\mu R}$ is the covariant derivative, with $A_{\mu L} = A_{\mu L}^a t ^a $ ($A_{\mu R} = A_{\mu R}^a t ^a $ ) denoting the gauge field corresponding to the left (right) $SU(N_f)$ group and $t^a$ the generators of the group. The field strength tensors are given by $F_{MN L,R} = \partial_M A_{N \ L,R}- \partial_N A_{M \ L,R} + g [A_{M \ L,R},A_{N \ L,R}]$, with the gauge choice in which $A_{5 L,R} = 0$ and $g_5$ denotes the five-dimensional gauge coupling identified using QCD Feynman diagram calculation of the Green's function as $g_5^2  = \frac{12 \pi^2}{N_c k}$. The bulk mass $M_5$ is given as $M_5^2= (\Delta - p)(\Delta + p - 4)k^2$, where $\Delta$ is the dimension of the $p-$form corresponding to the operator residing on the boundary~\cite{m5relation, GUBSER1998105}.

In the chiral action in Eq.\ref{eq12}, the bulk fields $X$, $A_{\mu L}^a$ and $A_{\mu R}^a$ correspond to the boundary operators $\bar{q}_Rq_L$, $\bar{q}_L \gamma^\mu t^a q_L$ and $\bar{q}_R \gamma^\mu t^a q_R$ respectively. Here $X$ is a bi-fundamental scalar field whose {\it vacuum expectation value} (vev) breaks the chiral symmetry spontaneously. From the covariant derivative term of the scalar field, the axial combination of $A_\mu = \frac{1}{2} (A_{\mu L}-A_{\mu R})$, gets a $z-$dependent mass term through the vev, while, the vector combination $V_\mu = \frac{1}{2} (A_{\mu L} +A_{\mu R})$, gets its mass from orbifold symmetry breaking. The holographic dual of the axial sector corresponds to the $a_1$ meson and its resonances, whereas the vector field is matched with the $\rho$ meson and its resonances.

\subsection{Scalar field}
We begin by analyzing the scalar sector. In the action presented above, the five-dimensional bulk scalar mass, with $\Delta =3$ and $p=0$, becomes,
\begin{equation}
    M_5^2= (\Delta)(\Delta- 4)k^2 = -3k^2 \ .
\end{equation}
Varying the action with respect to the scalar field, the equation of motion becomes,
\begin{equation}
    \Bigg[ 
        - \omega^3 \csc^3(k \omega z) \, \partial_\mu \partial^\mu 
        + \partial_z \Big( \omega^3 \csc^3(k w z) \, \partial_z \Big) 
        + 3k^2 \omega^5 \csc^5(k \omega z)
    \Bigg] X = 0 \ .
    \label{scalar_action}
\end{equation}
The solution to the above equation takes the form  $X(z) = \frac{1}{2}\left(c_1 F(z) + c_2 G(z)\right)$, where $F(z)$ and $G(z)$ are functions determined to be,
\begin{eqnarray}
\label{parametrisation}
        F(z) &=& \left( \frac{\csc^2\left(\frac{k \omega z}{2}\right)}{4 k \omega} 
    + \frac{\log\left( \cos\left(\frac{k \omega z}{2}\right)\right)}{k \omega} 
    - \frac{\log\left( \sin\left(\frac{k \omega z}{2}\right)\right)}{k w} 
    - \frac{\sec^2\left(\frac{k w z}{2}\right)}{4 k \omega} 
    \right) \sin^3(k \omega z) \nonumber \\
    G(z) &=& \frac{\sin^3(k \omega z)}{ (k \omega)^3} \ .
\end{eqnarray}
The integration constants $c_1$  and $c_2$ are parameterized, as $c_1 \rightarrow m_q k^{\frac{3}{2}}$  and $c_2 \rightarrow \sigma k^{\frac{3}{2}}$, respectively. For a vanishing induced cosmological constant, using Eqs.\ref{parametrisation}, the scalar field takes the form,
\begin{equation}
    X(z) \xrightarrow[]{\omega \rightarrow 0} \biggl( \frac{1}{2} m_q z + \frac{1}{2} \sigma z^3 \biggr) k^{3/2}
\end{equation}



\subsection{Vector meson part}
After spontaneous breaking of the scalar field $X$, the action in Eq.\ref{eq12} gives mass to the axial combination of $A_{\mu L}$ and $A_{\mu R}$. Thus, rewriting the gauge bosons $A_{\mu L}$ and $A_{\mu R}$ in the their mass basis as,
\begin{eqnarray}
     V_\mu &=& \frac{1}{2}(A_{\mu L} + A_{\mu R}) \nonumber \\
     A_\mu &=& \frac{1}{2}(A_{\mu L} - A_{\mu R}) \ .
\end{eqnarray}
The quadratic part of the gauge field becomes,
\begin{align}
    S_{\text{quad}} &= \int d^5x \left( -\frac{\omega \csc(k \omega z)}{8 g_5^2} \right) 
    \bigg[ 
        \partial_\mu V_\nu \partial^\mu V^\nu - \partial_\mu V_\nu \partial^\nu V^\mu 
        + \partial_\mu A_\nu \partial^\mu A^\nu - \partial_\mu A_\nu \partial^\nu A^\mu \nonumber \\
        &\quad - (\partial_\mu V_z - \partial_z V_\mu)^2 
        - (\partial_\mu A_z - \partial_z A_\mu)^2 
    \bigg] \ .
\end{align}
 To study the heavier spectrum and the corresponding mode functions of the vector and axial-vector mesons, it is enough to extract the quadratic terms in the action. While the boundary conditions on these fields at the UV brane are determined by the value of the sources of the vector and axial currents in four dimensions, at the IR brane, they are set to satisfy the Neumann condition.

The equation of motion for the transverse part of the vector field $V_\mu$, under the gauge $V_z(x,z) = 0$, becomes,
\begin{equation}
    \Bigg[\partial_z \Big(\omega \csc(k \omega z) \, \partial_z V_\mu(q, z)\Big) + q^2 \  \omega \csc(k \omega z) \ V_\mu(q, z)\Bigg]_\perp = 0 \ ,
    \label{eqn13}
\end{equation}
where $V_\mu^a (q,z)$ is obtained by a four-dimensional Fourier transform of $V_\mu^a (x,z)$. The analytical solutions to this equation are given by hypergeometric functions. Integrating the action over the extra dimension, the boundary term becomes,
\begin{equation}
    S = -\frac{1}{2g_5^2} \int d^4x \left(  \omega \csc(k \omega z) V_\mu \partial_z V^{\mu} \right)_{z=\epsilon} \ .
    \label{eqn14}
\end{equation}
Since the boundary condition at UV is obtained by matching the above with the source of the four-dimensional vector current, rewriting $V^{\mu }(q,z) = V(q,z)V_0^\mu(q) $  gives the condition $ V(q,\epsilon) = 1$. 

Varying the five-dimensional effective action with respect to the source, $V_0^\mu(q)$, the vector current two-point function becomes,
\begin{eqnarray}
        \int d^4x \, e^{iqx} \langle J_{\mu}^a(x) J_{\nu}^b(0) \rangle 
        &=& \delta^{ab} (q_{\mu} q_{\nu} - q^2 g_{\mu\nu}) \Pi_V(q^2) \ , \nonumber \\
        \Pi_V(q^2) &=& -\frac{\omega \csc(k \omega z)}{g_5^2 q^2} \partial_z V(q, z) \bigg|_{z = \epsilon}.
    \label{eqn15}
\end{eqnarray}

The solution of Eq.\ref{eqn13} is represented by $\psi_\rho(z)$ and is set to satisfy the boundary conditions, 
\begin{eqnarray}
    \psi_\rho(\epsilon) &=& 0 \ ,\nonumber \\
    \partial_z\psi_\rho(z_m) &=& 0 
    \label{bcs_rho}
\end{eqnarray}
and the normalization condition $\int dz \  \omega \csc(k \omega z) \psi_\rho(z)^2 = 1$. 

Identifying this solution with the hadrons requires writing the hadronic matrix element,
\begin{equation}
    \langle 0| J_\mu | \rho \rangle = F_\rho \epsilon_{\mu} \ ,
\end{equation}
where $\epsilon_\mu$ is the polarization of the $\rho$ meson and $F_\rho$ is the decay constant.
Using Eq.\ref{eqn15}, the $\rho$ meson decay constant can be related to the solution of Eq.\ref{eqn13} as~\cite{HadronModel},
\begin{equation}
    F_\rho^2 = \frac{1}{g_{5}^{2}} \left[ \omega \csc(k \omega z) \psi_\rho'(\epsilon) \right]^2\bigg|_{z=\epsilon}
    \label{eqn21}
\end{equation}

\subsection{Axial vector meson}
\label{sec:axialmeson}
In the axial sector, the action to quadratic order becomes (in the $A_z=0$ gauge),
\begin{equation}
S = \int d^5x \left[ -\frac{\omega \csc(k \omega z)}{4g_5^2}  F_{M N A} F^{M N}_A + \frac{v(z)^2 \omega^3 \csc^3(k \omega z)}{2} \left( \partial_M \pi (x) - A_{M} \right)^2 \right] \ .
\label{eqn22}
\end{equation}
The $z-$dependent vev of the scalar field $X$, $v(z)$, is found by solving Eq.\ref{scalar_action}. And $F_{MN A} $ is the field strength tensor of the axial vector $A_\mu = \frac{1}{2}(A_{\mu L} - A_{\mu R})$. The second term in the above action arises from the covariant derivative in Eq.\ref{scalar_action} where $\pi^at^a$ are the pseudo-Goldstone bosons arising from $X = v(z) \ e^{i2\pi^a (x) t^a}$. 

Upon separating the transverse and longitudinal parts in the axial vector field, $ A_{\mu}  =  A_{\mu \perp} + \partial_\mu \varphi$, and varying the action with respect to the transverse part of the field, we get the equation of motion,
\begin{equation}
\bigg[ \partial_z \bigg( \omega \csc(k \omega z) \partial_z \tilde{A}^a_\mu \bigg) 
+ q^2 \omega \csc(k \omega z) \tilde{A}^a_\mu 
- g_5^2 v(z)^2 \omega^3 \csc^3(k \omega z) \tilde{A}^a_\mu \bigg]_\perp = 0;
\label{eqn23}
\end{equation}
where $\tilde{A}^a_\mu$ is the Fourier transform of $A_{\mu \perp} $.
The above equation determines the physics of the bulk profile of the axial-vector meson $a_1$. Like in the case of the vector meson, the axil vector boson can be written as $\tilde{A}_{\mu \perp}^a (q,z) = \mathcal{A}(q,z)\tilde{A}_{\mu}^a(q)$ \cite{PhysRevD.76.115007}. Similarly, varying with respect to the longitudinal component gives,
\begin{eqnarray}
\partial_z \bigg( \omega \csc(k \omega z) \partial_z \varphi^a \bigg) 
+ g_5^2 v(z)^2 \omega^3 \csc^3(k \omega z) \big( \pi^a - \varphi^a \big) &=& 0 \ , \nonumber \\
-q^2 \partial_z \varphi^a + g_5^2 v(z)^2 \omega^2 \csc^2(k \omega z) \partial_z \pi^a &=& 0 \ .
\label{eqn25}
\end{eqnarray}
These equations can be solved numerically with the boundary conditions $\psi_{a_1}(\epsilon) = \partial_z\psi_{a_1}(z_m) = 0$, and $\varphi(\epsilon) = \partial_z\varphi(z_m) = \pi(\epsilon) = 0$, where $\psi_{a_1}$ is the wavefunction of the spin-1 axial meson. The $a_1$ decay constant, $F_{a_1}$, then becomes \cite{HadronModel},
\begin{equation}
    F_{a_1}^2 = \frac{1}{g_{5}^{2}} \left[ \omega \csc(k \omega z) \ \psi_{a_1}'(\epsilon) \right]^2\bigg|_{z=\epsilon}
    \label{fa1eqn}
\end{equation}

Since the axial-vector gauge symmetry is broken in five-dimensions via the vev of the $X$ field, the longitudinal component ($\partial_\mu \varphi^a$) is identified with the pseudo-Goldstone bosons, $\pi^a$. The pion wavefunction can be found by solving Eq.\ref{eqn25}, subject to the boundary conditions given above, assuming $q^2 = m_{\pi}^2$ with  $ \varphi(z) = \mathcal{A}(0,z) - 1 $, to the leading order of $m_\pi^2$ giving,
\begin{equation}
\pi(z) = \frac{m_{\pi}^2}{g_5^2} \int_{0}^{z} du \frac{\partial_u \mathcal{A}(0,u)}{v(u)^2 \omega^2 \csc^2(k \omega u)} \ .
\label{pimass}
\end{equation}
Here, $ \mathcal{A}(0, z) $ is the solution to Eq.\ref{eqn23} with $ q^2 = 0 $, satisfying axial boundary conditions $ \mathcal{A}'(0, z_m) = 0 $, $\mathcal{A}(0, \epsilon) = 1 $. Numerically, the integral gives $\pi(z) \rightarrow -1$, thus generating the Gell--Mann--Oakes--Renner relation,
\begin{equation}
  m_\pi^2 f_\pi^2 = 2 m_q \sigma + \mathcal{F}(z_m,\omega)  \ ,
  \label{eq:GORhardwall}
\end{equation}
where $\mathcal{F}$ is a function of the explicit symmetry breaking parameters $z_m$ and $\omega$. Note that, unlike the usual scenario discussed in literature~\cite{HadronModel}, here, $\omega$ brings in a new scale which has to be fitted phenomenologically.

Matching with the hadron current, $\langle 0|A_\mu|\pi \rangle = i f_\pi q_\mu$, the pion decay constant $f_{\pi}$ is identified as the coupling of the axial current in the limit $m_\pi = 0$. Like in the case of axial decay constant, the pion decay constant then becomes,
\begin{equation}
f_\pi^2 = -\frac{1}{g_5^2} \left. \omega csc(k \omega z) \partial_z \mathcal{A}(0,z) \right|_{z=\epsilon} \ .
\label{fpi}
\end{equation}

Below, we solve the equations of motion numerically and use the spectrum and the decay constants to compare between scenarios with vanishing and non-vanishing induced cosmological constant. Moreover, the error function in the GOR relation identified as $\mathcal{F}$ in Eq.\ref{eq:GORhardwall}, is fitted keeping $z_m$ and $\omega$ as free variables.

\subsection{Numerical results:}
\label{numerical}

\begin{figure}[h!]
    \centering
    \includegraphics[width=0.5\linewidth]{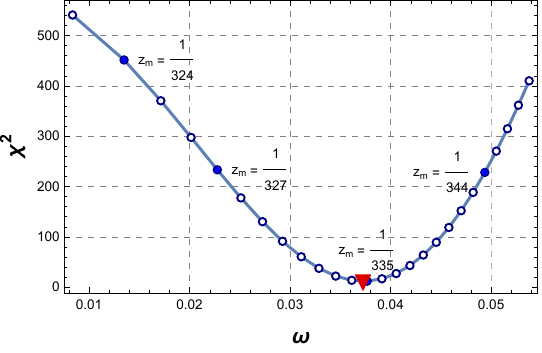}
    \caption{$\chi^2$ computed with the observables in Table~\ref{Model_Results} as a function of $\omega$, with minima $\approx 0.037$ with $z_m \approx 2.9 \text{GeV}^{-1}$. The minima is depicted as $(\textcolor{red}{\blacktriangledown})$.}
    \label{chisq}
\end{figure}

\begin{figure}[h!]
    \centering
    \includegraphics[width=0.5\linewidth]{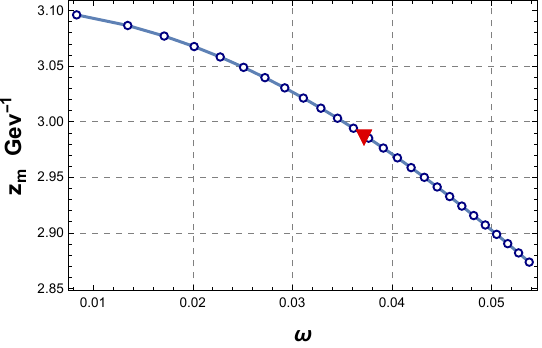}
    \caption{The figure shows the dependence of $z_m$ on $\omega$ after fixing $\rho$ meson mass in our model.}
    \label{wvszm}
\end{figure}

\begin{figure}[h!]
    \centering
    \begin{subfigure}{0.47\textwidth}
        \centering
        \includegraphics[width=\linewidth]{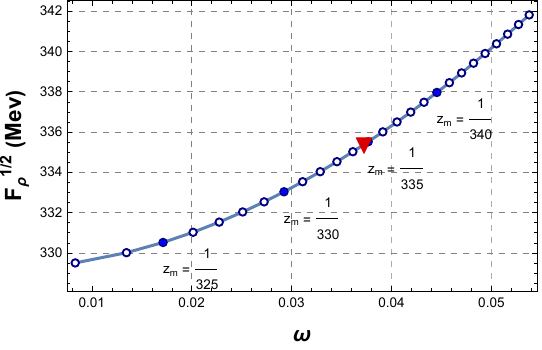}
        \caption{$\omega$ vs $F_\rho^{1/2}$}
        \label{wvsFrho}
    \end{subfigure} \hfill
    \begin{subfigure}{0.47\textwidth}
        \centering
        \includegraphics[width=\linewidth]{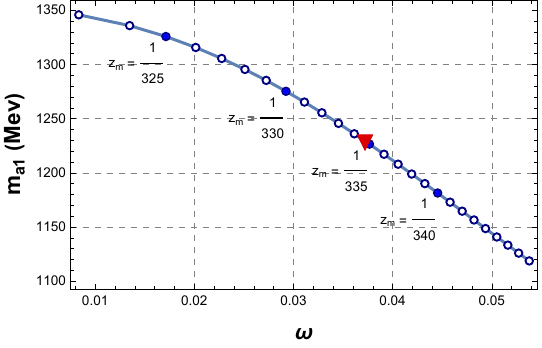}
        \caption{$w$ vs $m_{a1}$}
        \label{wvsma1}
    \end{subfigure} \hfill
    \begin{subfigure}{0.47\textwidth}
        \centering
        \includegraphics[width=\linewidth]{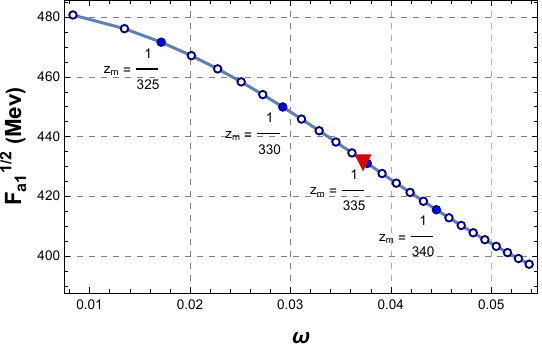}
        \caption{$\omega$ vs $F_{a1}^{1/2}$}
        \label{wvsFa1}
    \end{subfigure}
    \caption{Dependence axial-vector meson mass $m_{a1}$ and decay constants $F_{\rho}^{1/2}$, $F_{a1}^{1/2}$ on $\omega$.}
    \label{wvsobs}
\end{figure}

Here, we aim to find the model that best fits the experimental data given in the second column of Table \ref{Model_Results}. The model is determined by 5 parameters ($k, \omega, z_m, m_q, \sigma$), where $k$ is fixed by the bulk cosmological constant and five-dimensional Planck mass, which we set to 10 Gev. From the solution of the rho meson equation of motion, Eq.\ref{eqn13}, and using the boundary conditions Eq.\ref{bcs_rho}, we constrain the parameters $\omega$ and $z_m$ by fixing the $\rho$ meson mass to the experimental value $m_\rho = 775.26$ \cite{ParticleDataGroup}. Using numerical solution of Eq.\ref{eqn13}, and Eq.\ref{eqn23} we find the best fit to the observables $F_{\rho}^{1/2}$ $m_{a1}$, and $F_{a1}^{1/2}$ to find the dependence of $\chi^2$ on $\omega$, as shown in Fig.\ref{chisq}, for the values of $z_m$ already fitted with the $\rho$ mass. The best fit point is depicted as $(\textcolor{red}{\blacktriangledown})$ and the dependence of $z_m$ on $\omega$ is shown in Fig.\ref{wvszm}. And the individual dependence of observables $F_{\rho}^{1/2}$ $m_{a1}$, $F_{a1}^{1/2}$ on $\omega$ are shown in Fig.\ref{wvsFrho}, Fig.\ref{wvsma1}, and Fig.\ref{wvsFa1} respectively.

The minimum of the $\chi^2$ is the best fit of the model for the experimental errors, and corresponds to $\omega = 0.037$. The RMS value $\epsilon_{rms} = \sqrt{\frac{1}{n} \sum_{i=1}^{n} \left((O_{\text{exp},i} - O_{\text{model},i})/O_{\text{exp},i}\right)^2 }$ at the best fit for our model deviates from the experimental values at $1.6 \%$. In comparison, the model with vanishing $\omega$~\cite{HadronModel} incurs an error of $9.8 \%$. We can see that the model with $\omega = 0.037$ fits better with QCD.

\begin{figure}[h!]
    \centering
    \begin{subfigure}{0.47\textwidth}
        \centering
        \includegraphics[width=\linewidth]{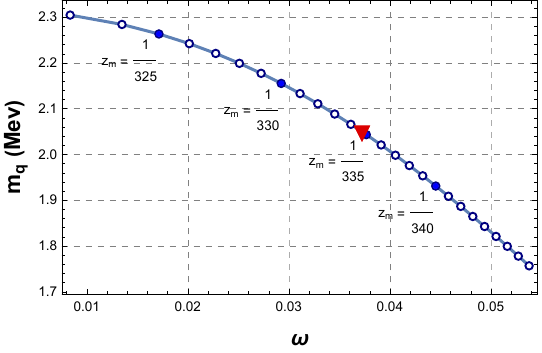}
        \caption{$\omega$ vs $m_q$}
        \label{wvsmq}
    \end{subfigure} \hfill
    \begin{subfigure}{0.47\textwidth}
        \centering
        \includegraphics[width=\linewidth]{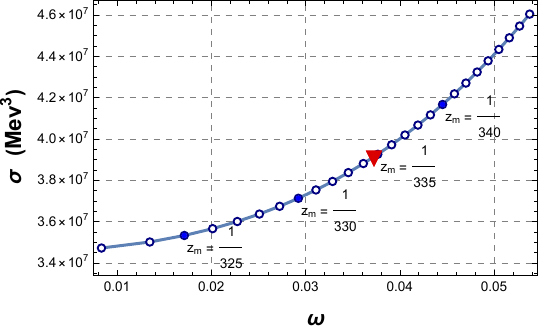}
        \caption{$w$ vs $\sigma$}
        \label{wvsS}
    \end{subfigure}
    \caption{The figure shows how the quark mass $m_q$ and the chiral condensate $\sigma$ as a function of $\omega$.}
    \label{w vs mq and s}
\end{figure}

\begin{figure}[h!]
    \centering
    \includegraphics[width=0.5\linewidth]{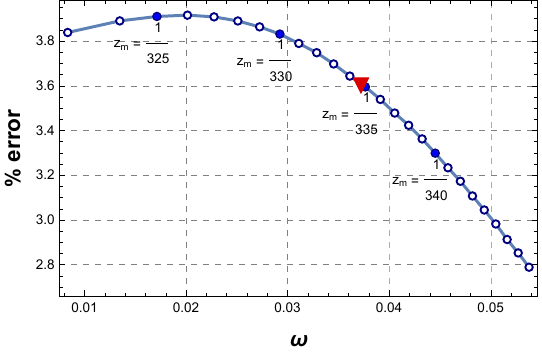}
    \caption{Dependence of the Gell-Mann-Oakes-Renner relation on $\omega$ is shown by taking the percent error. At the model best-fit point, we can see the relation is satisfied up to $3.6 \%$ accuracy.} 
    \label{fig:GORvsomegal}
\end{figure}

The remaining two model parameters, $m_q$ and $\sigma$ (whose dependence on $\omega$ can be seen in Fig.\ref{wvsmq} and \ref{wvsS} respectively), can be found by fitting the experimental values of $m_\pi$ and $f_\pi$ for given $\omega$. After fitting for $m_q$ and $\sigma$ for the minima of $\chi^2$, see that our model satisfies the Gell--Mann--Oakes--Renner relation, $ m_{\pi}^2 f_{\pi}^2 = 2m_q \sigma + \mathcal{F}(z_m,\omega)$ up to $3.6 \%$ accuracy for the best fit value $z_m=\frac{1}{335} \text{MeV}^{-1}$ and $\omega = 0.037$. The dependence of the error percentage in the GOR relation with $\omega$ is shown in Fig.\ref{fig:GORvsomegal}. Note that the error in the relation has an inverse dependence on the induced cosmological constant.

\begin{table}[H]
\centering
        \begin{tabular}{|c|c|c|c|}
            \hline
            \hline
            Observable  & Measured             &  $\omega=0$   & best fit $\omega_{min}=0.037$  \\
                            & (Mev) \cite{ParticleDataGroup}                      & (Mev) \cite{HadronModel}           & (Mev)            \\
            \hline
            \quad $m_\pi$         & $139.57 \pm 0.00018$& 139.57*             &  139.57*           \\
            \quad $m_\rho$         & $775.26 \pm 0.23$      & 775.26*           &  775.26*          \\
            \quad $m_{a_1}$       & $1230 \pm 40$              & 1363              &  $1228.8 \pm 5.3$ \\
            \quad $f_\pi$          & $92.4 \pm 0.35$           & 92.4*              &  92.4*            \\
            \quad $F_\rho^{1/2}$     & $345 \pm 8$ \cite{Donoghue_Golowich_Holstein_2014}  & 329                &  $335.39 \pm 0.27$         \\
            \quad $F_{a_1}^{1/2}$     & $433 \pm 13$ \cite{PhysRevD.69.065020}                & 486               &  $431.83 \pm 1.95$         \\
            \hline
            \hline
        \end{tabular}
        \captionof{table}{Comparison of Observables for the cases with and without $\omega$. Here, * denotes the observable values used to fit model parameters. The other three parameters are given with the $1-sigma$ error of $\chi^2$ Fig.\ref{chisq}. }
        \label{Model_Results}
\end{table}

\section{Soft wall with background dilaton field}
\label{dilaton}
\begin{figure}[h!]
    \centering
    \includegraphics[width=0.5\textwidth]{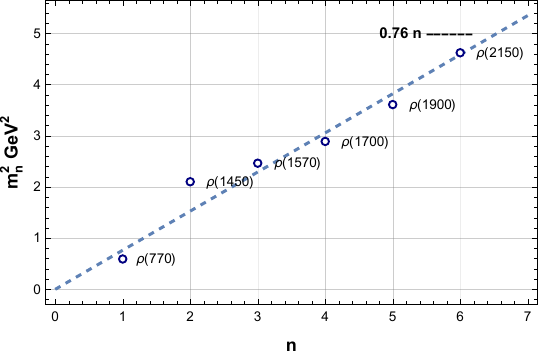}
    \caption{The squared masses of the first few $\rho$ resonances versus their consecutive number n. The dashed line shows the linear fit of the resonances taking $m_n^2 = \sigma n$}
    \label{nvrho}
\end{figure}

The drawback of this process so far is its inability to reproduce the higher excitations of vector and axial-vector mesons. For the hard wall scenario, where the large $z$ is cutoff by the IR brane, the mass squares ($m_n^2$) of the higher Kaluza-Klein modes depend on the corresponding Fourier level ($n$) as $m_n \sim n^2$. Identifying these Kaluza-Klein modes with the higher resonances of the mesons immediately brings in an issue. The experimental data on their spectrum suggests that the dependence goes as $m_n^2 \sim n$, as shown in Fig.\ref{nvrho}, and not the one predicted in the hard wall scenario. 

Fortunately, the behaviour of the Kaluza-Klein modes, in a warped extra-dimension scenario, is controlled by the IR rather than UV. Hence, the mass spectrum of these Fourier modes can be regulated by modifying the IR, keeping the UV asymptotically $AdS$. And introducing the dilaton contribution~\cite{LinearConfinement}, in the warped space-time, reproduced the Regge-like behaviour in the mass spectrum of the Kaluza-Klein modes. In the regular scenario, with vanishing induced cosmological constant, that would amount to the five-dimensional action in the background of a dilaton given by, 
\begin{equation}
    S=\int d^5x \sqrt{G}e^{-\Phi} \mathcal{L} \ ,
\end{equation}
where $\Phi$ is the dilaton field. To obtain $m_n^2\sim n$ spectrum, the $\Phi$ should grow as $ z^2$ in IR. Unlike the hard wall, here the $AdS$ space is segmented by a smooth cap. 

Here, the smooth IR wall (soft wall) is given by a gravitational background of the dilaton field $\Phi(z)$ and the metric $g_{M N}$. In the five-dimensional chiral Lagrangian, we also consider a cubic potential term for the scalar field given by \cite{PhysRevD.82.094013},
\begin{equation}
   S = \int d^5x \, e^{-\Phi(z)} \sqrt{G} \left\{ |DX|^2 + 3 k^2|X|^2 + \frac{\lambda k^2}{\sqrt{2}} |X|^3 - \frac{1}{4g_5^2}(F_L^2 + F_R^2)  \right\}
\label{dilatonaction}
\end{equation}
The bulk scalar {\it vacuum expectation value} ($v(z)$) and the background dilaton $\Phi(z)$ are related by the equation of motion derived from the bulk action Eq.\ref{dilatonaction},
\begin{equation}
    \partial_z \Big( \omega^3 \csc^3(k w z) \, e^{-\Phi(z)} \, \partial_z v\Big)
    + \omega^5 \csc^5(k \omega z) \, e^{-\Phi(z)} \Big(3k^2 v + \frac{3}{4} \lambda k^{1/2}\ v^2  \Big)
    = 0 \ .
    \label{eq:vev}
\end{equation}
Writing this in terms of dilaton $\Phi$~\cite{PhysRevD.82.094013, Gherghetta:2009ac} we get,
\begin{equation}
    \Phi'(z) = -3k \omega \cot(k \omega z) + 
\frac{
  3 \sqrt{k} \omega^2 \csc^2(k \omega z) \, v(z) \left(4 k^{3/2} + \lambda v(z) \right) + 4 v''(z)
  }{4 v'(z)} \ .
    \label{dilatonDeqn}
\end{equation}
And the boundary conditions become, 
\begin{eqnarray}
    v &\sim& \left( m_q \zeta z + \frac{\sigma}{\zeta} z^3 \right)k^{3/2}  \ \ \ \  z \rightarrow \epsilon  \nonumber \\
    v &\sim& z^2 \ \ \ \ z \rightarrow \text{IR} \ ,
\end{eqnarray}
 where the constant $\zeta = \sqrt{3}/(2\pi)$\cite{PhysRevC.79.045203}. The scaling $k^{3/2}$ comes from the parametrisation used in Eq.\ref{parametrisation}. Thus, ansatz to the solution for Eq.\ref{dilatonDeqn} that satisfy the above boundary conditions can be found as~\cite{PhysRevD.82.094013},
\begin{equation}
    \frac{1}{2} v(z) = \frac{A z + B z^3}{\sqrt{1 + C z^2}}\, k^{3/2} .
\end{equation}
We aim to study the effect of induced cosmological constant, $\omega$, in the model parameters $A, B, C,\lambda$. The relation between $m_q $ and $\sigma$ with $A, B, C$ become,
\begin{align}
    m_q = \frac{2 A}{\zeta}, \quad \quad \sigma = 2 \zeta(B - \frac{1}{2} A C^2).
    \label{mqandswithdilatonparameters}
\end{align}

Varying the action in Eq.\ref{dilatonaction} with the vector boson, as we did in the hard wall scenario, the equation of motion becomes,
\begin{equation}
    \partial_z \Big(\omega \csc(k \omega z) \, e^{-\Phi(z)} \, \partial_z V_\mu^n\Big) + q^2 \  \omega \csc(k \omega z) \, e^{-\Phi(z)} \ V_\mu^n = 0 \ ,
    \label{rhodilaton}
\end{equation}
where $V_n$ are the Kaluza-Klein modes of the vector boson. Identifying the KK modes with the higher resonances of the $\rho$ meson, the mass spectrum is obtained from normalizable solutions of the Eq.\ref{rhodilaton} which satisfy the boundary conditions,
\begin{eqnarray}
    V_\mu^n(\epsilon) &=& 0 \nonumber \\
    \partial_z V_\mu^n\Big(\frac{\pi}{2 k \omega}\Big) &=& 0 \ ,
\end{eqnarray}
where $\frac{\pi}{2 k \omega}$, gives the natural cut-off of the warped space-time, given in Eq.\ref{ads}. The mass spectrum we obtain is discussed in the next section. For $\omega = 0.005$, the best-fit point of the model, the spectrum is given in Table.\ref{dilatonrhotable}.

Similarly, to obtain the spectra of axial vector mesons, we vary the action Eq.\ref{dilatonaction}, as was done in Sec.\ref{sec:axialmeson}, and expanding it in terms of its KK modes $A_\mu^n$, the equation of motion becomes,
\begin{equation}
\partial_z \bigg( \omega \csc(k \omega z) \, e^{-\Phi(z)} \partial_z A_{\mu}^n \bigg) 
+ q^2 \omega \csc(k \omega z) \, e^{-\Phi(z)} A_{\mu}^n
- g_5^2 v(z)^2 \omega^3 \csc^3(k \omega z) \, e^{-\Phi(z)} A_{\mu}^n= 0;
\label{a1dilaton}
\end{equation}
The effective four-dimensional action for the tower of massive axial-vector fields can be identified as the $a_1$ mesons together with their radial excitation states, satisfying the boundary conditions,
\begin{eqnarray}
    A_\mu^n(\epsilon) &=& 0 \nonumber \\
    \partial_z A_\mu^n\Big(\frac{\pi}{2 k w}\Big) &=& 0 \ .
\end{eqnarray}
Again, the spectrum of axial vector mesons is discussed in the next section after fitting the model with the experimental data. For the best fit of the model, $\omega = 0.005$, the spectrum is given in Table.\ref{dilatona1table}.

\subsection{Numerical results:}
\label{dilatonresults}
\begin{figure}[H]
    \centering
    \includegraphics[width=0.5\linewidth]{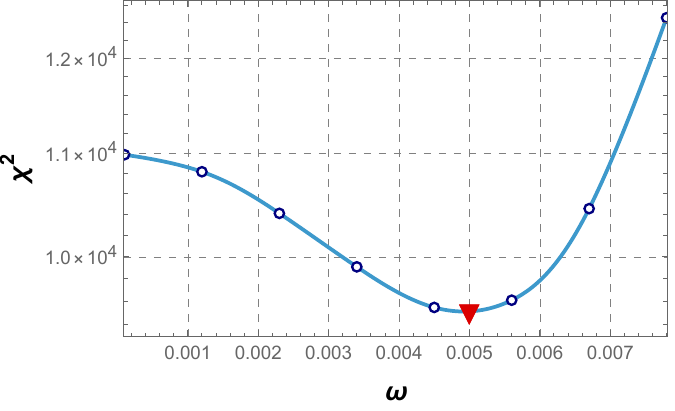}
    \caption{$\chi^2$ vs $\omega$ plot with the minima at $\omega_{min} \approx 0.005$, highlighted as $(\textcolor{red}{\blacktriangledown})$ for the chosen value of model parameters.} 
    \label{chisqdilaton}
\end{figure}

The model parameters $A, B, C,$ and $ \lambda $ are determined by fitting to the experimental data, specifically, the masses of the $\rho$, $ a_1 $ mesons and their heavier modes, as well as the pion mass $ m_{\pi} $ and the pion decay constant $ f_{\pi} $. We use the total relative error square, $\mathscr{E}^2 = \sum_{i} \left( \frac{O_i^{\mathrm{theory}} -O_i^{\mathrm{exp}}}{O_i^{\mathrm{exp}}} \right)^2$, which is minimized, to fit the model parameters. The values for these model parameters after fitting become,
\begin{equation}
    A = 0.51\, \text{MeV},\quad B = (360.9\, \text{MeV})^3,\quad C = 2234.74\, \text{MeV},\quad \lambda = 25.28 \ .
    \label{dilaton_parameter_values}
\end{equation}
Then, using the Eq.~\ref{mqandswithdilatonparameters} we obtain the corresponding values of $m_q$ and $\sigma$ as,
\begin{equation}
    m_q = 3.7\ \text{Mev}, \quad \sigma = (293.24\ \text{Mev})^3 \ .
\end{equation}
To determine the optimal value of $\omega$ that best fits the model, we perform a $\chi^2$ analysis incorporating the relevant observables : $m_\pi$, $f_\pi$, $F_\rho^{1/2}$, $F_{a_1}^{1/2}$ given in Table. \ref{Model_Results}, along with the mass spectra of the $\rho$ and $a_1$ mesons in Table.~\ref{dilatonrhotable} and Table.~\ref{dilatona1table}. The dependence of $\chi^2$ on the induced cosmological constant, $\omega$, is shown in Fig.\ref{chisqdilaton}. 

\begin{figure}[h!]
    \centering
    \includegraphics[width=0.5\linewidth]{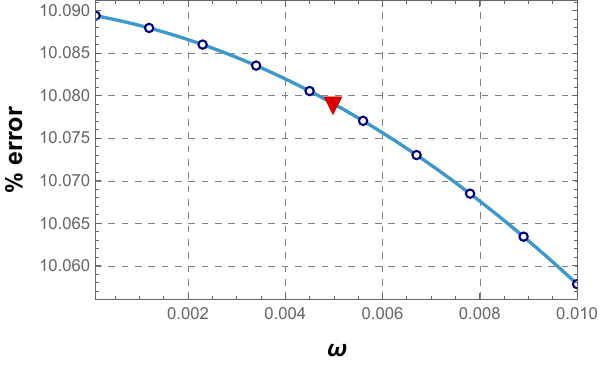}
    \caption{Dependence of Gell--Mann--Oakes--Renner relation on $\omega$.}
    \label{fig:DilatonGORvsomega}
\end{figure}

We see that the spectra of $\rho$ and $a_1$ mesons give us asymptotically linear spectra, and the model results with/without $\omega$ are listed in Tables. \ref{dilatonrhotable} and \ref{dilatona1table}. Computing the GOR relation, with the best fit model parameters, we again observe that our model satisfies the relation up to better accuracy compared to the model with vanishing induced cosmological constant, as shown in Fig.\ref{fig:DilatonGORvsomega}. Though the $\omega$ dependence of error, in the soft wall model, is not stark like in the hard wall scenario, nevertheless, the behaviour is consistent. The $AdS_5/QCD$ error in GOR relation has an inverse relation with the induced cosmological constant. 
\begin{table}[H]
    \centering
    \begin{tabular}{|c||c|c|c|}
    \hline
    \hline
         n &   $\rho$ experimental       & $\rho$ Model (MeV)  & $\rho$ Model (MeV) \\
           &          (MeV)              & $\omega = 0 $       & $\omega_{min} \approx 0.005$\\
    \hline
         1 &      775.26 $\pm$ 0.23      &     754.87          &        760.27      \\
         2 &      1282   $\pm$ 37        &     1173.33         &        1189      \\
         3 &      1465   $\pm$ 25        &     1477.09         &        1505.73      \\
         4 &      1720   $\pm$ 20        &     1727.92         &        1771.58      \\
         5 &      1909   $\pm$ 30        &     1946.69         &        2006.89      \\
         6 &      2149   $\pm$ 17        &     2145.23         &        2221.67      \\
         7 &      2265   $\pm$ 40        &     2335.8          &        2423.21      \\         
    \hline
    \hline
    \end{tabular}
    \caption{Spectra of $\rho$ mesons and its higher resonances.}
    \label{dilatonrhotable}
\end{table}
\begin{table}[H]
    \centering
    \begin{tabular}{|c||c|c|c|}
    \hline
    \hline
         n &   $a_1$ experimental        & $a_1$ Model (MeV)  &$a_1$ Model (MeV)\\
           &         (MeV)               & $\omega = 0 $       & $\omega_{min} \approx 0.005$\\
    \hline
         1 &      1230   $\pm$ 40        &      1009.16        &     1011.78         \\
         2 &      1647   $\pm$ 22        &      1537.57        &     1545.77         \\
         3 &      $1930^{+ 30}_{-70}$    &      1924.77        &     1940.05         \\
         4 &      2096   $\pm$ 122       &      2245.56        &     2269.15         \\
         5 &      $2270^{+ 55}_{-40}$    &      2525.63        &     2558.58         \\      
    \hline
    \hline
    \end{tabular}
    \caption{Spectra of $a_1$ mesons and its higher resonances.}
    \label{dilatona1table}
\end{table}

\section{Summary}
\label{summary}

In contrast to standard approaches that assume an AdS background with Minkowski metric induced on the hypersurfaces, here, we consider a generalized gravitational setup with an induced cosmological constant. By embedding the chiral Lagrangian into this generalized gravitational background, we aim to explore the sensitivity of the predicted hadronic properties and their spectra to changes in the underlying spacetime curvature, thereby providing new insights into the holographic realization of symmetry breaking.

In the article, we study the effect of the induced cosmological constant in a five-dimensional holographic model. Using the five-dimensional chiral Lagrangian and its symmetry-breaking pattern in a hard wall $AdS_5/QCD$ model, we compare our results with measured QCD meson observables. We also examine how these observables vary with the induced cosmological constant, parameterized by $\omega$ as shown in Fig.\ref{wvszm} and Fig.\ref{wvsobs}. A $\chi^2$ analysis, shown in Fig.\ref{chisq}, reveals that our best-fit model, with an induced cosmological constant, achieves a lower root-mean-square (RMS) error of 1.6\%, compared to 9.8\% in the limit $\omega \to 0$. Moreover, the percentage error in the Gell-Mann-Oakes--Rener relation reduces, see Fig.\ref{fig:GORvsomegal}, hence preferring an $AdS_5/QCD$ model with non-vanishing induced cosmological constant. 

Furthermore, in Section \ref{dilaton}, we introduce a background dilaton field within a soft-wall holographic model to predict higher meson resonances. Performing a $\chi^2$ analysis over all relevant observables $m_\pi, f_\pi, F_\rho^{1/2}, F_{a_1}^{1/2}$, and mass spectra of $\rho$ and $a_1$ mesons, we again find that the best-fit model corresponds to a $AdS_5$ geometry with non-vanishing value for the induced cosmological constant. We consistently observe an inverse correlation between the percentage error in the GOR relation and $\omega$.

\section*{Acknowledgments}
M.T.A. acknowledges the financial support of DST through the INSPIRE Faculty grant  DST/INSPIRE/04/2019/002507. The Authors are grateful to Giancarlo D`Ambrosio for valuable discussions.

\nocite{*}
\bibliographystyle{unsrt} 

\bibliography{ref} 

\end{document}